# Nickel (II) Ions Interaction with Polynucleotides and DNA of Different GC Composition

*Vasil G. Bregadze[1], Irina G.Khutsishvili[1], Sophie Z. Melikishvili[1], Zaza G. Melikishvili[2]*

[1]Andronikashvili Institute of Physics, 6 Tamarashvili Street, Tbilisi 0177, Georgia

[2]Insttute of Cybernetics, 5 Sandro Euli Street, Tbilisi 0186, Georgia

**Abstract.** The goal of the work was to study the role of GC alternative dimmers in the binding of DNA with Ni (II) ions. The method of ultraviolet difference spectroscopy has been applied to investigate Ni (II) ions interactions with DNA extracted from *Clostridium perfringens*, Mice liver (C3HA line), Calf thymus, Salmon sperm, Herring sperm, *E.coli*, *Micrococcus luteus* and polynucleotides Poly (dA-dT)×Poly (dA-dT), Poly (dG)×Poly (dC), Poly (dG-dC)×poly (dG-dC). It is shown that Ni (II) ions at outer-spherical binding with DNA double helix from the side of the major groove choose more stable dimmers

$$\begin{array}{c} 3' - C - G - 5' \\ \cdot \quad \cdot \\ 5' - G - C - 3' \end{array}$$

and get bound with $N_7$ atoms of both guanines in dimmer forming G-G interstrand crosslink. It directly correlates to the process of forming point defects of Watson-Crick wrong pair type_(creation of rare keto-enolic and amino-imino tautomeric forms) and depurinization.

**Abbreviations:** UVS, ultraviolet spectra; UDS, ultraviolet difference spectra; EB, ethidium bromide; $\Delta\varepsilon_s = [\Delta\varepsilon_{max}] + [\Delta\varepsilon_{min}]$, summery values of molar coefficient of extinction; M(II), metal (II).

## 1. Introduction

Due to recently discovered DNA catalytic properties interest in inorganic biochemistry has increased. It was discovered that, DNA double helix possesses mediator properties in photodynamic affect [1-4], as well as ones connected with charge [5-12] and energy [13-15] transfers. In all those processes transitive metal ions and their complexes with small organic molecules, for example metalointercalators [1, 16, 17] and metal based drugs [18-25], play important if not major role. Besides some transitive metal ions and their compounds are strong cancerogens and mutagens. The mutations that lead to death of the cells are initiated by antitumor drugs. Often such drugs contain metal ions. Sigels' new series "Metal Ions in Life

Sciences," that has been published recently contains complete review connected with complex formation of Ni (II) and related metal ions with nucleic acids and with its components [26]. Though in our present paper we want to draw attention to special feature Ni (II) ions at its interacting with DNA, as it gives preference to stable GC alternating dimmers. For this purpose, interaction of Ni (II) ions with polynucleotides and DNA of different origin was investigated by the method of ultraviolet difference spectra.

The reason for choosing Ni (II) ions are the following:

1. NI (II) ions as well as their complexes are strong cancerogens [27-31].
2. Ni (II) ions are strong agents which in small quantities, in certain conditions B $\rightarrow$ Z conformational transition in DNA double helix [32].
3. Some first-transition-row ions like Mn (II), Fe (II), Co (II), Ni (II) and Zn (II) form the outer-spherical complex with DNA nitrogen bases. Ni (II) is most active from them [14].

Earlier the interaction of $H_3O^+$ ions, transitive metals and intercalators with DNA was studied by the thermodynamic and spectroscopic methods. Suggested thermodynamic model of ions-DNA interaction establishes direct proportional connection between the dynamic characteristic of the interaction i.e. lifetime of the complex and its equilibrium characteristics stability constant. It is shown that from electron-donor-atoms of B-DNA double helix $N_7G$, $N_3G$, $N_3A$, $O_2T$, $O_6G$ and $N_7A$ nitrogen bases, only $N_7G$ has the advantage to participate in outer-spherical interaction of hexaaqua ions with DNA [14,33]. The role of $H_3O^+$ and TM in keto-enolic and amino-imino tautomeric transitions in DNA base pairs and depurination was studied. The probabilities and energies of rare tautemeric forms of GC pairs in DNA induced by $H_3O^+$ and TM were determined [34]. The present paper is devoted to investigation of GC alternative dimmers role in Ni (II) -DNA interaction.

## 2. Materials and Methods

### Materials

**DNA and Polinucleotides:** In the paper we used the DNA from calf thymus (40% GC) [35], herring sperm (44% GC) [35] *micrococcus luteus* (72% GC) [36], *clostridium perfringens* (27% GC) [36], salmon sperm (44% GC) [35] , *E. coli* (51% GC) ("sigma"), and liver of C3HA line mice (40% GC) [35] extracted by Dr. N. Sapozhnikova in the Andronikashvili Institute of Physics. Besides, we used polynucleotides Poly (dG-dC) ×Poly (dG-dC), Poly (dG) ×Poly (dC) and Poly (dA-dT) ×Poly (dA-dT) (firm "Boehringer Mannheim").Concentration of DNA and polynucleotides were also determined spectrophotometrically and as molar coefficients ε for these polymers we have taken the following values: *E. coli*-6500, calf thymus-6600, *micrococcus luteus*-6420, *clostridium perfringens*-6700, salmon sperm-6500, Poly (dG-dC) ×Poly (dG-dC) and Poly (dG) ×Poly (dC)-7100.

**Ions:** We used chemically pure chloride of Ni and also especially pure NaCl. Bidistillate served as a solvent. The concentration of the filtered mother liquor of NI (II) was determined spectrophotometrically.

**Intercalator:** Ethidium bromide we also purchased from "Sigma". The concentration of the dye was determined colorimetrically (5600 at 480 nm).

**Method**

Ultra-violet difference spectra (UDS) of DNA caused by its interaction with two-charged metal ions were registered by double-beam spectrometer Specord M40 (firm Carl Zeiss) in carefully matched 1cm long quartz cells. Detailed description of the procedure is given in [14].

## 3. Results and Discussions

Fig.1 presents UDS of DNA from the cell thymus, *micrococcus luteus*, *clostridium perfringens* and also Poly (dG-dC) ×Poly (dG-dC) and Poly (dG) ×Poly (dC), caused by their interaction with NI (II) ions. The analysis of their spectra has shown that despite some differences in the position of UDS maxima in the region of 290 nm; on the whole the spectra coincide qualitatively but differ significantly in the intensity of Ni (II) ions influence on ultra-violet spectra of these polymers.

Fig.2 presents UDS of DNA caused by its interaction with Ni (II) ions and ethidium bromide. More detailed results of investigation of the complexes of Ni (II) with DNA, with various nucleotide compositions are given in the table 1.

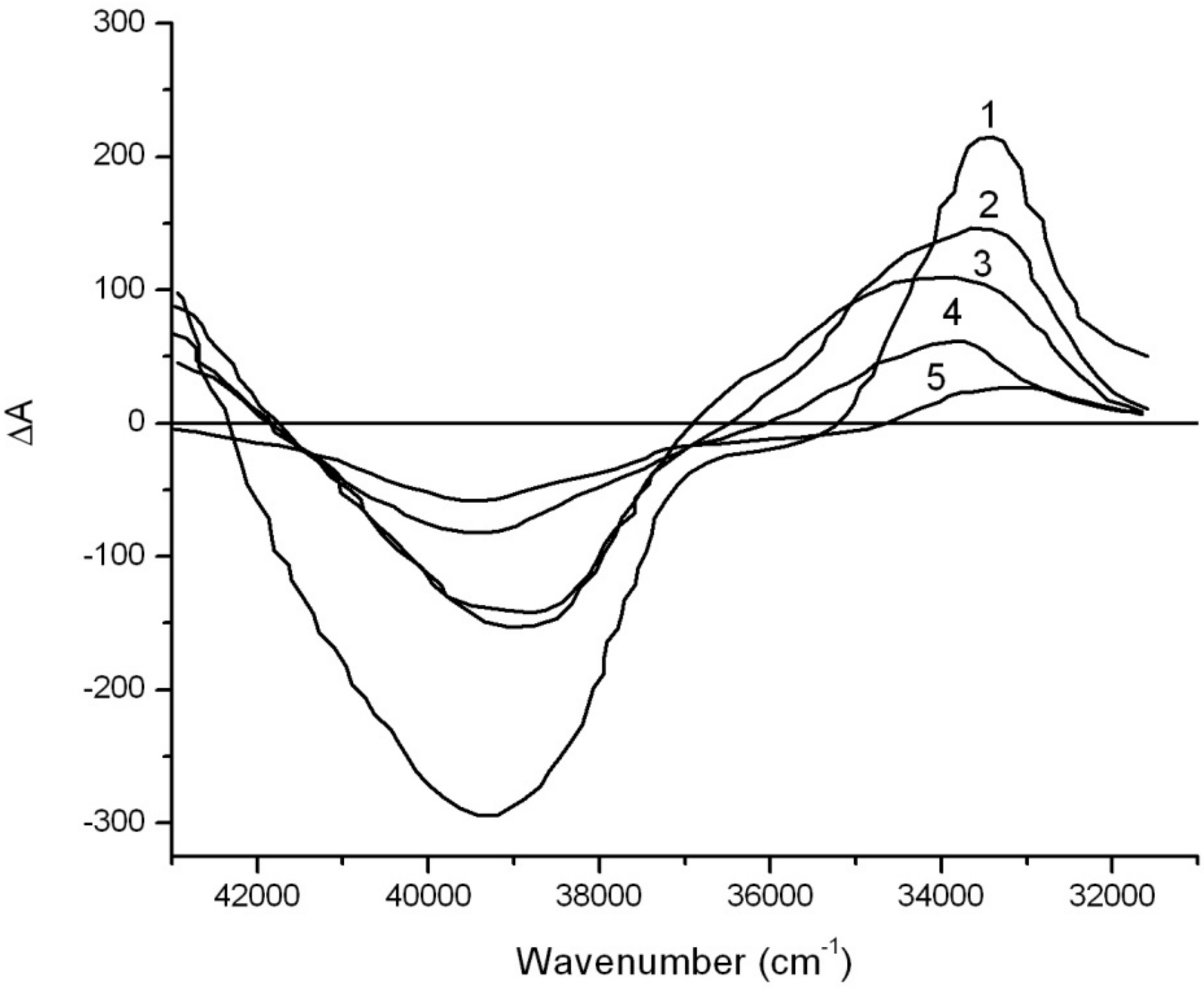

**Fig.1** Ultra-Violet difference spectra between complexes: 1 – Poly (dG-dC) ×Poly (dG-dC); 2 – DNA from *micrococcus luteus*; 3 – DNA from calf thymus; 4 – Poly (dG) ×Poly (dC); 5 – DNA from *clostridium perfringens* with Ni (II) ions and polymers in the absence of the ion. Concentrations: DNA and polynucleotides - 2× $10^{-4}$ M by phosphorus; NaCl - $10^{-2}$ M; Ni (II) - 0.25 per polymer nucleotide.

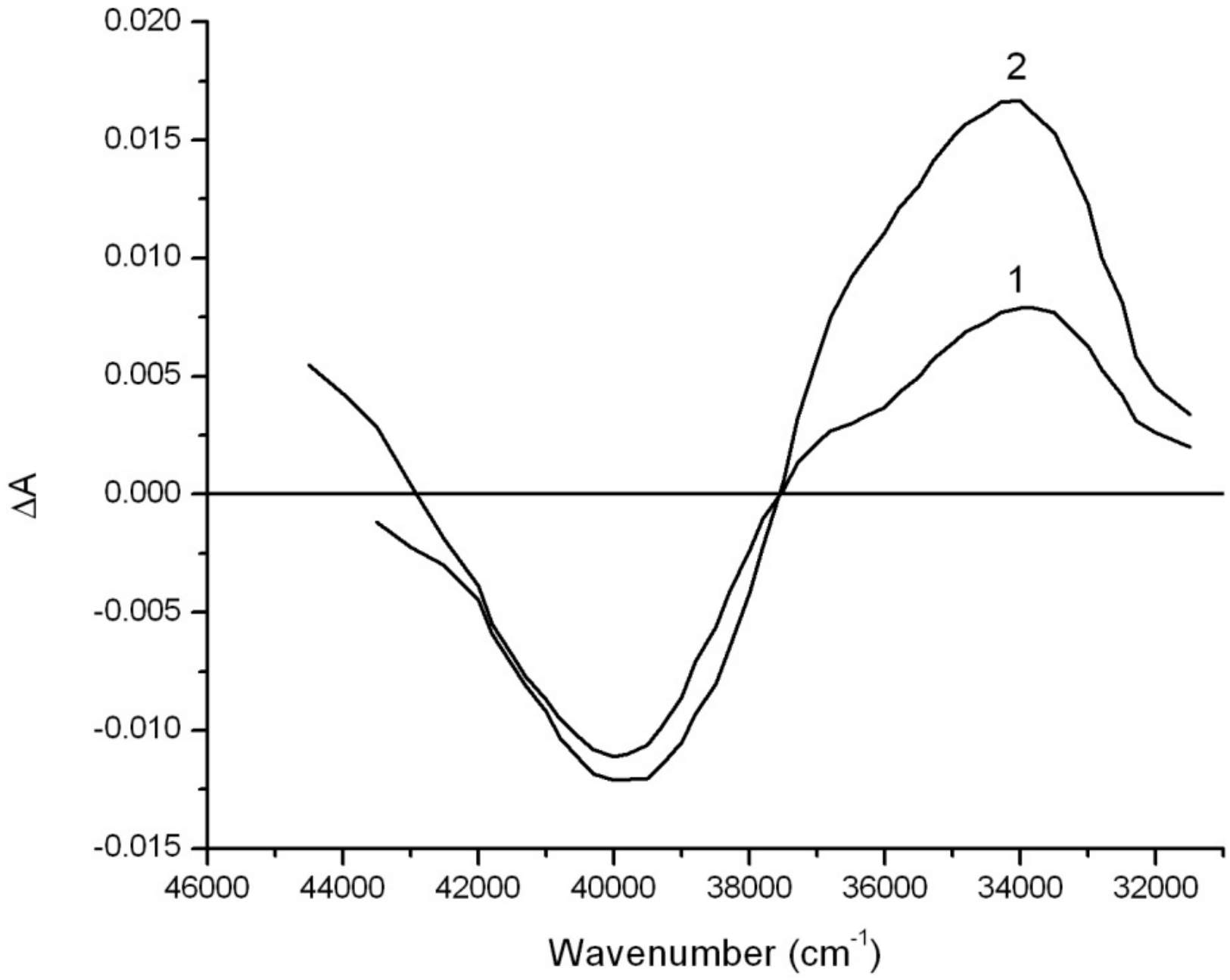

**Fig. 2** DNA UDS with Ni (II) ions and EB. 1-UDS between DNA complexes with NI (II) ions and DNA in the absence of the ion; 2-UDS between DNA complexes with EB and Ni (II) ions and DNA complexes with EB but in the absence of Ni (II) ions. Concentrations: DNA- $10^{-4}$ M (P); EB- $10^{-5}$ M; $NiCl_2$- 2.5×$10^{-4}$ M; NaCl- $10^{-2}$ M.

**Table 1.**

UDS characteristics of NI (II) ions complexes (0.25 Ni (II)/P) with polynucleotides and DNA of various nucleotide composition.

| # | **Polynucleotides and DNA from** | **%GC** | **$\Delta\varepsilon_s$** | **$E_{i/uvs}$** |
|---|---|---|---|---|
| 1 | Poly (dA-dT) Poly (dA-dT) | 0 | 0 | - |
| 2 | Clostridium perfringens | 27 | 100 | 3.7 |
| 3 | Mice liver (C3HA line) | 40 | 180 | 4.5 |
| 4 | Calf thymus | 40 | 230 | 5.75 |
| 5 | Calf thymus + EB (0.1 EB/ P) | 40 | 275 | 6.9 |
| 6 | Salmon sperm | 44 | 260 | 5.9 |
| 7 | Herring sperm | 44 | 295 | 6.7 |
| 8 | E. coli | 51 | 360 | 7.1 |
| 9 | Micrococcus luteus | 72 | 335 | 4.7 |
| 10 | Poly (dG) Poly (dC) | 100 | 145 | 1.5 |
| 11 | Poly (dG-dC) Poly (dG-dC) | 100 | 530 | 5.3 |

In the first column the sources of DNA extraction and polynucleotides used during our work are presented. The second column shows the percent of GC composition in the investigated samples. In the third column we have $\Delta\varepsilon_s = [\Delta\varepsilon_{max}] + [\Delta\varepsilon_{min}]$ - summery values of molar coefficient of extinction [14]. The last column shows values of efficiency of Ni (II) ions influence on DNA UV absorption spectra, $E_{i/uvs} = \Delta\varepsilon/\%GC$ relation of $\Delta\varepsilon_s$ to percent GC composition in the samples under investigation. All these values $\Delta\varepsilon$ are given for relative concentration of Ni (II) equal to 0.25 per nucleotide. The last row of the table demonstrates the results of EB affect on Ni (II) – DNA interaction.

Analysis of dates given in the table 1, as well as analysis of dates given by Fig.1 and Fig.2 shows that Ni (II) ions interacting with DNA double helix give preference to GC alternating dimmers. Those dimmers are found in two types:

1) $\begin{matrix} 3' - C - G - 5' \\ \cdot \quad \cdot \\ 5' - G - C - 3' \end{matrix}$ 2) $\begin{matrix} 3' - G - C - 5' \\ \cdot \quad \cdot \\ 5' - C - G - 3' \end{matrix}$

Their stacking energies are 14.59 and 9.69 kcal per mole of dimmer correspondingly [37]. It must be mentioned that intercalators, including EB are located between base pairs, formed by self-complimentary dinucleoside monophosphates, which always have pyrimidine nucleoside on the 5' end and purine nucleoside on the 3'end [38]. I.e. intercalators give preference to GC dimmers of

$$\begin{matrix} 3' - \mathrm{G} - \mathrm{C} - 5' \\ \quad\cdot \quad \cdot \\ 5' - \mathrm{C} - \mathrm{G} - 3' \end{matrix}$$

second type [39-41]. It is interesting that intercalation in this case is energetically more preferable by 7-13 kcal/mol than in the case of dimmer of first type [42]. Ultraviolet difference spectra given on the Fig. 2 clearly demonstrate that Ni (II) ions and EB molecules have different binding sites. In addition the values of stability constant depend on double helix dynamical properties and structure stability. EB gives preference to the following dimmers

$$\begin{matrix} 3' - \mathrm{G} - \mathrm{C} - 5' \\ \quad\cdot \quad \cdot \\ 5' - \mathrm{C} - \mathrm{G} - 3' \end{matrix}$$

while Ni (II) ions give preference to dimmers

$$\begin{matrix} 3' - \mathrm{C} - \mathrm{G} - 5' \\ \quad\cdot \quad \cdot \\ 5' - \mathrm{G} - \mathrm{C} - 3' \end{matrix}$$

On intercalation unwinds double helix approximately on 18°and it involves at least three base pairs at each side of the intercalation site.Besedes,. EB also elongate the double helix and increases its hardness [43]. In addition Ni (II) ions notably increase DNA thermostability [44].

Therefore, we can conclude that hexaaqua ions of Ni (II), or at list their certain quantity, bind form outer-spherically G-C dimmers of the first type in double helix from the side of the major groove thus forming bridges between $N_7G$ of one chain and $N_7G$ of the opposite chain

$$\begin{matrix} 3' - \mathrm{C} - \mathrm{G} - 5' \\ \quad\cdot \quad \cdot \\ 5' - \mathrm{G} - \mathrm{C} - 3' \end{matrix}$$

It must be mentioned that interaction between ions of transitive metals, including Ni (II) ions, with DNA is very diverse. It depends on double helix structure, nucleotide sequences, ionic force, dielectric transmission of solvent, as well as on ion type, its electronic structure, and on diversity of aqua complexes. The molecule of water really has two nonbonding electron pairs on the $sp^3$ -hybrid orbital of oxygen atom. The distance between their centers is 0.6 Å [45]. Certainly, the pairs are equivalent, but the orientation of the tetrahedron depends on the pair participating in electron-donor interaction. Thus, the hexaaqua ion may be in $2^6$ different states, if we neglect the interaction of water molecules in the complex. During the time interval exceeding the time of water molecule stay in the first hydrate layer of M (II) ion ($10^{-4}$-$10^{-7}$ s), the situation will be averaged, but in each time moment the molecule of water will be turned to the ion of the electron pairs.

It should be noted that though the binding of the ions of the first transitive row with DNA has diverse character (innerspherical, outerspherical, doubleouterspherical, atmospherical, with participation of phosphate group and nitrogen bases), chelatic interaction of Ni(II) ions with the atoms of both guanines in GC dimmers forming G-C interstrand crosslink is of special importance for molecular biology as it is in direct correlation with the process of forming of point defects of Watson- Crick wrong pair type (creation of rare keto-enolic and amino-imino tautomeric forms) and depurinization [14, 46, 34, 23].

We should also point out that GC alternative polynucleotide

$$\begin{array}{c} 3' - C - G - C - G - 5' \\ \cdot \quad \cdot \quad \cdot \quad \cdot \\ 5' - G - C - G - C - 3' \end{array}$$

has 0.5 M of dimmers (types I and II) per 1 M of GC b.p. If we consider that Ni (II) ions can interact exclusively with type I dimmers then due to strong Coulomb repulsion between Ni(II) ions the number of places for outer-spherical interaction from the side of major groove is not more than 0.25M. On the other hand, if we imagine that polynucleotide has GC and AT alternative dimmers

$$\begin{array}{c} 3' - C - G - T - A - 5' \\ \cdot \quad \cdot \quad \cdot \quad \cdot \\ 5' - G - C - A - T - 3' \end{array}$$

then the number of Ni(II) binding places with both nucleotides from the side of major groove will be the same (0.25 M), while the efficiency of the influence on DNA UV spectra $E_{i/uvs}$ is twice as big for the second nucleotide and is 10.6 instead of 5.3 (see Table 1). This value greatly exceeds the data for consequent efficiency observed at investigated DNAs. Increase or decrease of $E_{i/uvs}$ value compared to the date given in Table for poly (dG-dC)-poly (dG-dC)(5.3) proves the uniqueness of nucleotide sequence in DNA. Thus, out of 64 triplets of gene code 14 ones contain GC dimmers 8 of which are of type I. 4 of 8 dimmers can encode amino acid alanine and the other single ones encode arginine, serine, cysteine and glycine, correspondingly. Besides, there can be a situation when each of neighboring triplets is not so active at interaction with Ni (II) ions, but performing together they become quite active as in the case given below.

RNA $$5' - \overbrace{U - U - G}^{\text{Leu}} - \overbrace{C - A - U}^{\text{His}} - 3'$$

DNA
$$\begin{array}{c} 3' - A - A - C - G - T - A - 5' \\ \cdot \quad \cdot \quad \cdot \quad \cdot \quad \cdot \quad \cdot \\ 5' - T - T - G - C - A - T - 3' \end{array}$$

So based on the interaction principals of Ni (II) ions with DNA, it is shown that the sequence of nucleotides in DNA plays major role in this process, especially presence of GC stable dimmers of $A - G$

$$\begin{array}{c} 3' - C - G - 5' \\ \cdot \quad \cdot \\ 5' - G - C - 3' \end{array}$$

type.

**Acknowledgements**

The authors express deep gratitude to Mrs Greta Nijaradze for help in preparing the manuscript.